\documentclass[twocolumn,showpacs,preprintnumbers,amsmath,amssymb]{revtex4}
\usepackage{graphicx}
\usepackage{dcolumn}
\usepackage{bm}

\begin{document}

\preprint{APS/123-QED}

\title{Generalized ABCD propagation for interacting atomic clouds.}

\author{F. Impens$^{1,2}$ and Ch. J. Bord\'{e}$^{1,3}$}

\affiliation{$^{1}$ SYRTE, Observatoire de Paris, CNRS, 61 avenue
de l'Observatoire, 75014 Paris, France}

\affiliation{$^{2}$ Instituto de Fisica, Universidade Federal do
Rio de Janeiro. Caixa Postal 68528, 21941-972 Rio de Janeiro, RJ,
Brasil}

\affiliation{$^{3}$ Laboratoire de Physique des Lasers, Institut
Galil\'{e}e,  F-93430 Villetaneuse, France}

\date{\today}

\begin{abstract}
We present a treatment of the nonlinear matter wave propagation
inspired from optical methods, which includes interaction effects
within the atom optics equivalent of the aberrationless
approximation. The atom-optical ABCD matrix formalism, considered
so far for non-interacting clouds, is extended perturbatively
beyond the linear regime of propagation. This approach, applied to
discuss the stability of a matter-wave resonator involving a
free-falling sample, agrees very well with the predictions of the
full nonlinear paraxial wave equation. An alternative optical
treatment of interaction effects, based on the aberrationless
approximation and suitable for cylindrical paraxial beams of
uniform linear density, is also adapted for matter waves.
\end{abstract}


\maketitle

\section{INTRODUCTION}

Light and matter fields are governed by similar equations of
motion~\cite{BordeHouches}. Both photons and atoms interact in a
symmetrical manner: atom-atom interactions are mediated through
photons, while photon-photon interactions are mediated through
atoms. Before the advent of Bose-Einstein condensation, two groups
realized independently that atomic interactions give rise to a cubic
nonlinearity in the propagation equation analogous to that induced
by the Kerr effect~\cite{Lenz93,Zhang94}. Following this analogy,
the field of non-linear atom optics emerged in the last decade,
leading to the experimental verification with matter waves of
several well-known nonlinear optical phenomena\footnotemark[1]
\footnotetext[1]{Many other optical phenomena have also been
verified with matter waves. A short list includes
interferences~\cite{BermanBook} and diffraction
phenomena~\cite{Zhang94,Krutitsky}, the temporal Talbot
effect~\cite{Deng99}, and the influence of spatial phase
fluctuations on interferometry~\cite{Jo07}. New effects arise also
with rotating condensates~\cite{Josopait03}.}: the four-wave
mixing~\cite{DengScience99}, the formation of
solitons~\cite{Burger99,Denschlag00,Salomon02,Strecker02} and of
vortices~\cite{Matthews99,Madison00}, the
superradiance~\cite{Inouye99superrad} and the coherent
amplification~\cite{Ketterle99}. The nonlinear propagation of matter
waves has been the object of extensive
experimental~\cite{LeCoq01,Busch02} and theoretical work, among
which the time-dependent Thomas-Fermi
approximation~\cite{CastinDum96}, the variational
approach~\cite{Michinel95}, and the method of
moments~\cite{Moments}. These treatments have been used successfully
to obtain analytical expressions in good agreement with the exact
solution of the 3D nonlinear Schr\"odinger equation
(NLSE).\\

There exists, for cylindrical wave-packets propagating in the
paraxial regime, a very elegant method to handle this equation
which has been used in optics to treat self-focusing
effects~\cite{Yariv78,Belanger83}. It relies on the
``aberrationless approximation'', assuming that the nonlinearity
is sufficiently weak as to preserve the shape of a fundamental
Gaussian mode, and it involves a generalized complex radius of
curvature. This treatment is equally relevant for the paraxial
propagation of cylindrical matter waves, and it is  presented in
this context in Appendix~\ref{app:belanger}. Unfortunately, the
assumptions required - such as the constant longitudinal velocity
and the paraxial propagation - limit the scope of this approach,
which appears as too stringent to describe the matter wave
propagation in most experiments.\\

This motivates the introduction of a different analytical method to
obtain approximate solutions for the NLSE in a more general
propagation regime. This is the central contribution of this paper,
which exposes a perturbative matrix analysis especially well-suited
to discuss the stability of a matter-wave resonator. With an
 Hamiltonian quadratic in position and momentum operators, and in the absence of atomic interactions, the
Schr\"odinger equation admits a basis of Gaussian solutions. Their
evolution is easily obtained through a time-dependent matrix
denoted ``ABCD''~\cite{BordeHouches,BordeMetrologia2002}, in
analogy with the propagation of optical rays in
optics~\cite{Kogelnik65}. In the ``aberrationless approximation'',
it is possible to extend this treatment to include perturbatively
interaction effects and obtain the propagation of a fundamental
Gaussian mode with a modified ``ABCD'' matrix. As an illustration
of this method, the stability of a matter-wave resonator is
analyzed thanks to this ``ABCD'' matrix, which encapsulates the
divergence resulting from the mean-field potential. An ABCD-matrix
approach had already been used in~\cite{LeCoq01} to characterize
the divergence of a weakly outcoupled atom laser beam due to
interactions with the source condensate. The present treatment is
sensibly different, since it is not restricted to the paraxial
regime and since it addresses rather self-interaction effects in
the beam propagation. An ``ABCD'' matrix, including self-focusing
effects, is computed in Sec.~\ref{sec:ABCDmatrix}, and used to
model the propagation of an atomic sample in a matter-wave
resonator. Self-focusing is also discussed through an alternative
method exposed in Appendix~\ref{app:belanger}.\\

Our approach is indeed mainly inspired from previous theoretical
developments in optics, which aimed at treating the wave
propagation in a Kerr medium through such a matrix
formalism~\cite{Yariv78}. An approach of the non-linearity based
on the resulting frequency-dependent diffraction~\cite{Garside68}
successfully explained the asymmetric profile of atomic and
molecular intra-cavity resonances~\cite{LeFloch80}, as well as the
dynamics of Gaussian modes in ring and two-isotopes
lasers~\cite{Bretenaker90a,Bretenaker90b}. Later, a second-order
polynomial determined by a least-square fit of the wave intensity
profile was considered to model the Kerr effect~\cite{Magni93}. In
this paper, we explore the quantum mechanical counterpart of this
strategy: mean-field interactions are modelled thanks to a
second-order polynomial, determined perturbatively from the
wave-function, and which can be interpreted in optical terms.

\section{LENSING POTENTIAL}
\label{sec:lensing effect}

 One considers the propagation of a
zero-temperature condensate in a uniform gravity field and in the
mean-field approximation. The corresponding Hamiltonian reads:
\begin{equation}
\label{eq:basic Hamiltonian} \hat{H}=  \frac {\hat{\mathbf{p}}^2}
{2m} + m g z+  g_{I}  |\phi(\hat{\mathbf{r}},t)|^2
\end{equation}
$g_{I}$ is the coupling constant related to the s-wave scattering
length $a$ and to the number of atoms $N$ by $g_{I}= 4 \pi N \hbar^2
a / {m}$. Our purpose is to approximate the mean-field potential $
g_{I} |\phi(\mathbf{r},t)|^2$ by an operator leading to an easily
solvable wave equation and as close as possible to the interaction
potential. A second-order polynomial
 in the position and momentum operators is a suitable choice, since it allows to obtain
Gaussian solutions to the propagation equation. These solutions
are approximate, but they lead nonetheless to a satisfactory
description of the propagation of diluted atomic wave-packets and
of their stability in resonators, which are the issues addressed
in this paper. We note $\hat{H}_0=\frac {\hat{\mathbf{p}}^2} {2m}
+ m g z$ the interaction-free Hamiltonian and
\begin{equation}
\label{eq:definition hamiltonian function u}
\hat{H}(\hat{\mathbf{r}}, \hat{\mathbf{p}},t)=\hat{H}_0+P_l(
\hat{\mathbf{r}}, \hat{\mathbf{p}},t)
\end{equation}
the quadratic Hamiltonian accounting for interactions effects.\\

The strategy exposed in this paper consists in picking up, among
the possible polynomials $P$, the element which minimizes an
appropriate distance measure to the mean-field potential. In
geometric terms, this polynomial appears as the projection of the
mean-field potential onto the vector space spanned by second-order
polynomials in position and momentum. This potential will be
referred to as the ``lensing potential'', denomination which will
be justified in Sec. \ref{sec:ABCDmatrix}. We define a distance
analogous to the error function used in~\cite{Magni93}, which
involves the polynomial $P$ and the quantum state $|\phi(t)
\rangle$ resulting from the non-linear evolution:
\small
\begin{equation}
\label{eq:error function} E  \left(   P(t)  ,  |\phi(t) \rangle
\right)  =  \int d^3\mathbf{r} \left| \langle \mathbf{r} | P(
\hat{\mathbf{r}},\hat{\mathbf{p}},t) | \phi(t) \rangle -
g_{I}|\phi(\mathbf{r},t)|^2 \phi(\mathbf{r},t) \right|^2
\end{equation}
\normalsize
 The minimization of the distance $E \left(  \:  P(t) \: , \: |\phi(t) \rangle \:  \right)$ for the
lensing potential $P_l(t)$ implies that the function  $E$ is
stationary towards any second-order polynomial coefficient
 at the point $(P_l(t),|\phi(t) \rangle)$:
\begin{equation}
\label{eq:equation fondamental vecteur u} \forall t \: \geq \: t_0
\quad \quad \: \nabla_{P} \:  E \: (  \: P_l(t) \:, \: |\phi(t)
\rangle \: ) \: =  \: 0
\end{equation}
We have noted $\nabla_{P}$ the gradient associated with the
coefficients of a second-order polynomial, and $t_0$ is the initial
time from which we compute the evolution of the wave-function - we
assume that $\phi(\mathbf{r},t_0)$ is known -. The determination of
the lensing potential associated with self-interactions in the beam
indeed requires previous knowledge of the wave-function evolution.
This difficulty did not arise in other optical treatments of atomic
interaction effects~\cite{LeCoq01,Riou06,Riou08}, in which the
atomic beam propagation was mainly affected by interactions with a
different sample of well-known wave-function. This is typically
 the case for a weakly outcoupled continuous atom laser beam, in which the diverging lens effect
 results from the source condensate. We propose to circumvent this self-determination problem thanks
 to a perturbative treatment. Such approach is legitimate for the diluted matter waves involved in usual
atom interferometers. The first-order lensing polynomial and the
corresponding Hamiltonian
$\hat{H}^{(1)}(t)=\hat{H}_0+P_l^{(1)}(\hat{\mathbf{r}},
\hat{\mathbf{p}},t)$ are determined from the linear evolution,
according to:
\begin{equation}
\label{eq:definition u1} \forall t \: \geq \: t_0 \quad \quad
\nabla_{P} \: E \left(P_l^{(1)}(t) \: , \: e^{- i / \hbar
\hat{H}_0 (t-t_0)} |\phi(t_0) \rangle \right) =  0
\end{equation}
Higher-order lensing effects can be computed iteratively. For
instance, the second-order lensing polynomial $P_l^{(2)}(t)$
satisfies at any instant $t \: \geq \: t_0$:
\begin{equation}
 \nabla_{P}  E  \left(P_l^{(2)}(t),
T\left[ e^{- i / \hbar \int_{t_0}^{t} dt'
[\hat{H}_0+P_l^{(1)}(\hat{\mathbf{r}}, \hat{\mathbf{p}},t)]}
\right] |\phi(t_0) \rangle \right)  =  0 \nonumber
\end{equation}
where we have used the usual time-ordering operator $T$~\cite{Peskin}.

\section{OPTICAL PROPAGATION OF MATTER WAVES: THE ABCD THEOREM.}

This section gives a remainder on a general result - called the ABCD
theorem -  concerning the propagation of matter waves in a
time-dependent quadratic potential, which is the atomic counterpart
of the ray matrix formalism frequently used in
optics~\cite{Kogelnik65}. It shows that the evolution of a Gaussian
wave-function under an Hamiltonian quadratic in position and
momentum is similar to the propagation of
 a Gaussian mode of the electric field in a linear optical system. A detailed
 description of this theoretical result of atom optics is given in the references~\cite{BordeMetrologia2002,BordeTheortool2001}.\\

One considers a time-dependent quadratic Hamiltonian such as:
\begin{eqnarray}
\label{eq:introduction hamiltonien quadratique general} \hat{H}_0
& + & P_l( \hat{\mathbf{r}}, \hat{\mathbf{p}},t)  =  \frac{
\hat{\widetilde{\mathbf{p}}} \beta(t) \hat{\mathbf{p}}} {2 m} +
\frac {1} {2} \hat{\widetilde{\mathbf{p}}} \alpha(t)
\hat{\mathbf{r}} -  \frac {1} {2} \hat{\widetilde{\mathbf{r}}}
\delta(t) \hat{\mathbf{p}} \nonumber
\\ & - & \frac {m} {2} \hat{\widetilde{\mathbf{r}}} \gamma(t)
\hat{\mathbf{r}} - m \mathbf{g}(t) \cdot \hat{\mathbf{r}}
 +\mathbf{f}(t) \cdot \hat{\mathbf{p}}+h(t)
\end{eqnarray}
$\alpha(t)$, $\beta(t)$, $\gamma(t)$ and $\delta(t)$ are $3 \times
3$ matrices \footnotemark[2]
\footnotetext[2]{$\delta(t)=-\widetilde{\alpha}(t)$ to ensure the
Hamiltonian hermiticity}; $\mathbf{f}(t)$ and $\mathbf{g}(t)$ are
three-dimensional vectors; $h(t)$ is a scalar, and $\widetilde{}$
stands for the transposition. Here we use this Hamiltonian to
approximate the nonlinear Hamiltonian~\eqref{eq:basic
Hamiltonian}. The Hamiltonian~\eqref{eq:introduction hamiltonien
quadratique general} is indeed appropriate to describe several
physical effects~\cite{BordeDirac00,BordeGRG2004}.

\subsection{ABCD propagation of a Gaussian wave-function.}
\label{eq:ABCD theorem}

The propagation of a Gaussian wave-packet in such an Hamiltonian
can be described simply as follows. Let $\phi(\mathbf{r},t)$ be an
atomic wave packet initially given by:
\begin{equation}
\label{eq:fonction initiale gaussienne} \phi (\mathbf{r},t_0) =
\frac { 1} {\sqrt{|\mbox{det} X_0|}} e^{\frac {i m} {2 \hbar}
(\mathbf{r}-\mathbf{r}_{c0}) Y_0 X_0^{-1}
(\mathbf{r}-\mathbf{r}_{c0}) + \frac i \hbar \mathbf{p}_{c0} \cdot
(\mathbf{r}-\mathbf{r}_{c0})}
\end{equation}
The $3 \times 3$ complex matrices  $X_0$, $Y_0$ represent the
initial width of the wave packet in position and momentum
respectively: $X_0= i\mbox{D}(\Delta x(t_0), \Delta y(t_0), \Delta
z(t_0))$ and $Y_0= \mbox{D}(\Delta p_x(t_0), \Delta p_y(t_0), \Delta
p_z(t_0))$, with $\mbox{D}$ standing for a diagonal matrix. The
vectors $\mathbf{r}_{c0}$, $\mathbf{p}_{c0}$ give the initial
average position and momentum. The ABCD theorem for matter waves
states that, at any time $t \geq t_0$, the wave-packet $\phi
(\mathbf{r},t)$ satisfies:
\begin{equation}
\phi (\mathbf{r},t)= \frac {e^{\frac i \hbar
S(t,t_0,\mathbf{r}_{c0},\mathbf{p}_{c0})}} {\sqrt{|\mbox{det} X_t|}}
e^{\frac {i m} {2 \hbar} (\mathbf{r}-\mathbf{r}_{c  t}) Y_t X_t^{-1}
(\mathbf{r}-\mathbf{r}_{c t}) + \frac i \hbar \mathbf{p}_{c t} \cdot
(\mathbf{r}-\mathbf{r}_{c t})} \nonumber
\end{equation}
$S(t,t_0,\mathbf{r}_{c0},\mathbf{p}_{c0})$ is the classical action
evaluated between $t$ and $t_0$ of a point-like particle which
motion follows the classical Hamiltonian
$H(\mathbf{r},\mathbf{p},t)$  and with respective initial position
and momentum $\mathbf{r}_{c0},\mathbf{p}_{c0}$. The width matrices
in position $X_t$ and momentum $Y_t$, and the average position and
momentum $\mathbf{r}_{c t},\mathbf{p}_{c t}$ at time $t$ are
determined through the same $6 \times 6$  ``ABCD'' matrix:
\begin{eqnarray}
\label{eq:introduction relation ABCD position et impulsion
moyenne}
\left( \begin{array} {c}\mathbf{r}_{c t} \\
\frac {1}  {m}  \mathbf{p}_{c t}\end{array} \right) & = &
 \left(
\begin{array}{cc}
A(t,t_0) & B(t,t_0)  \\
C(t,t_0) & D(t,t_0)  \\
\end{array} \right)
\left( \begin{array} {c}\mathbf{r}_{c0} \\
\frac {1}  {m}  \mathbf{p}_{c0} \end{array} \right) + \left( \begin{array} {c} \mathbf{\xi}(t,t_0) \\
 \mathbf{\phi}(t,t_0) \end{array} \right) \nonumber \\
\left( \begin{array} {c}X_t \\
Y_t\end{array} \right) & = &
 \left(
\begin{array}{cc}
A(t,t_0) & B(t,t_0)  \\
C(t,t_0) & D(t,t_0)  \\
\end{array} \right)
\left( \begin{array} {c} X_0 \\
Y_0\end{array} \right) \nonumber
\end{eqnarray}
The ABCD matrix -noted compactly $M(t,t_0)$-  and the vectors
$\mathbf{\xi},
 \mathbf{\phi}$ can be expressed formally as~\cite{BordeGRG2004}:
 \small
\begin{eqnarray}
\label{eq:expression formelle matrice ABCD}
 M(t,t_0)& = & T \left[ \exp \left(  \int_{t_0}^{t}
dt' \left(
\begin{array}{cc}
\alpha(t') & \beta(t')  \\
\gamma(t') & \delta(t')  \\
\end{array} \right) \right) \right]  \\
\label{eq:expression formelle vecteur xi phi}
 \left( \begin{array} {c} \mathbf{\xi}(t,t_0) \\
 \mathbf{\phi}(t,t_0) \end{array} \right) & = &  \int_{t_0}^t dt' \: M(t,t') \: \left( \begin{array} {c} \mathbf{f}(t') \\
 \mathbf{g}(t') \end{array} \right)
\end{eqnarray}
\normalsize Although the former expressions seem rather involved,
in all cases of practical interest, the $ABCD\xi \phi$ parameters
can be determined analytically or at least by efficient numerical
methods.

\subsection{Interpretation of the ABCD propagation and aberrationless approximation.}

The phase-space propagation provides a relevant insight in the
transformation operated by the ABCD matrix. Consider the Wigner
distribution of a single-particle density operator evolving under
the Hamiltonian \eqref{eq:introduction hamiltonien quadratique
general}. The Wigner distribution at time $t$ is related to the
distribution at time $t_0$ by the following map:
\begin{eqnarray}
W \left( \mathbf{r},\mathbf{p},t \right) & = & W \left( \frac {}
{} \right. \widetilde{D} (\mathbf{r} - \mathbf{\xi}) - \frac 1 m
\widetilde{B} (\mathbf{p} - m \mathbf{\phi}) \: ,  \nonumber \\
& \: & \left. \quad \quad  - m \widetilde{C} (\mathbf{r} -
\mathbf{\xi}) + \widetilde{A} (\mathbf{p} - m \mathbf{\phi}) \: ,
\: t_0 \frac {} {} \right) \nonumber
\end{eqnarray}
where the matrices $A,B,C,D$ and vectors $\xi, \phi$ are again
evaluated at the couple of instants $(t,t_0)$. The action of the
evolution operator onto the Wigner distribution is thus amenable to
a time-dependent linear map. The fact that ABCD matrices are
symplectic~\cite{BordeMetrologia2002} implies that this map is
unitary: such evolution preserves the global phase-space volume, and
the quality factor of an atomic beam in
the sense of~\cite{Impens08a}.\\

In photon as in atom optics, the aberrationless approximation
consists in assuming that the Gaussian function~\eqref{eq:fonction
initiale gaussienne} is a self-similar solution of the propagation
equation in spite of the non-linearity, the evolution of which is
given by the ABCD propagation. The propagation is thus described
through a map which preserves the phase-space density. This is an
approximation, since for atomic or light beams evolving in
nonlinear media, the phase-space density indeed changes during the
propagation. Nonetheless, the aberration-less approximation is
reasonable for sufficiently diluted clouds, subject to a weak
mean-field interaction term, for which an initially Gaussian
wave-function will not couple significantly to higher-order modes.
Furthermore, this approximation in atom optics is entirely
analogous to the aberration-free treatment realized in non-linear
optics, the predictions of which concerning the width evolution of
a light beam have been verified experimentally~\cite{Nemoto95}.
One can thus expect that the aberrationless approximation will
constitute a good description of the propagation in atom optics as
well. Indeed, the validity of the aberrationless approximation
will be confirmed in Sec.~\ref{subsec:comparison aberationless
paraxial} on the example of a gravitational atomic resonator: its
predictions on the sample size evolution are in good agreement
with those of a paraxial treatment of the wave-function
propagation which does not assume the preservation of a Gaussian
shape.


\section{ABCD MATRIX OF A FREE-FALLING INTERACTING ATOMIC CLOUD.}
\label{sec:ABCDmatrix}

Let us apply the method discussed above to describe the propagation
of a free-falling Gaussian atomic wave-packet. In the aberrationless
approximation, such a wave-packet is simply determined by the
parameters $ABCD\xi\phi$  and by the phase associated with the
action. In view of the resonator stability analysis, we will focus
on the computation of the ABCD matrix in presence of the mean-field
potential. We consider only the leading-order nonlinear corrections,
associated
 with the first-order lensing polynomial
 $P_l^{(1)}(\mathbf{r},\mathbf{p},t)$.\\

  This section begins with the determination of this potential defined by Eq.\eqref{eq:definition u1}. A formal expression
  of the atom-optical ABCD matrix, taking into account this
  lensing
  potential, is obtained. An infinitesimal expansion of this expression shows that the mean-field interactions effectively play the role of a divergent
  lens: the atom-optical ABCD matrix of the free-falling cloud evolution is similar to the optical ABCD matrix associated with the propagation
   of a light ray through a series of
infinitesimal divergent lenses. In our case, the propagation axis
is the time, and the infinitesimal lenses correspond to the action
of the mean-field potential in infinitesimal time slices.

\subsection{Determination of the lensing potential.}
\label{subsec:determination lensing pragraph} We assume that the
condensate, evolving in the Hamiltonian~\eqref{eq:basic
Hamiltonian}, is initially at rest and described by a Gaussian
wave-function:
\begin{equation}
\label{eq:Gaussian wave function} \phi(x,y,z,t_0)= \frac
{\pi^{-3/4}} { \sqrt{w_{x 0} w_{y 0} w_{z 0}}} e^{- \frac {x^2} {2
w_{x 0}^2}-\frac {y^2} {2 w_{y 0}^2}-\frac { z^2} {2 w_{z 0}^2}}
\end{equation}
It is easy to show that, when one considers the interaction-free
evolution, the widths are given at time $t \geq t_0$ by:
\begin{equation}
\label{eq:linear width evolution} w_{i t}= \sqrt{w_{i 0} ^2+\frac
{\hbar^2} {m^2 w_{i 0}^2} (t-t_0)^2}
\end{equation}
for $i=x,y,z$. This result can be easily retrieved by considering
the initial width matrices $X_0= i \mbox{D}(  w_{x 0}, w_{y 0}, w_{z
0})$ and $Y_0= \frac {\hbar} {m} \mbox{D} (1/w_{x 0}, 1 / w_{ y 0},
1 / w_{z 0})$ for the wave-function, and applying the free ABCD
matrix~\cite{BordeMetrologia2002}:
\begin{equation}
 \left(
\begin{array}{cc}
A(t,t_0) & B(t,t_0)  \\
C(t,t_0) & D(t,t_0)  \\
\end{array} \right)= \left( \begin{array} {cc} 1 & t-t_0 \\
0 & 1 \end{array} \right) \nonumber
\end{equation}
The square of the free-evolving wave-function thus reads:
\begin{equation}
|\phi^{(0)}(\mathbf{r},t)|^2=  \frac {\pi^{-3/2}} {w_{x t} w_{y t}
w_{z t}} e^{- \frac {(x-x_{c t})^2} { w_{x t}^2}-\frac {(y-y_{c
t})^2} {w_{y t}^2}-\frac {(z-z_{c t})^2} {w_{z t}^2}} \nonumber
\end{equation}
We use this expression to determine the first-order lensing
polynomial $P^{(1)}_l(\mathbf{r},\mathbf{p},t)$. Since this
operator acts on Gaussian wave-functions, differentiation is
equivalent to the multiplication by a position coordinate, so the
action of the momentum operator is indeed equivalent to that of
the position operator up to a multiplicative constant. One can
thus, without any loss of generality, search for a lensing
polynomial $P^{(1)}_l(\mathbf{r},t)$ involving only the position
operator. With this choice, the error function~\eqref{eq:error
function} minimized by the polynomial $P$ becomes simply:
\small
\begin{equation}
\label{eq:lensing polynomial error function} E( P(t) \: , \:
|\phi^{(0)}(t) \rangle) = \int d^{3} \mathbf{r}
|\phi^{(0)}(\mathbf{r},t)|^2 \left( P(\mathbf{r},t) -
   g_{I} |\phi^{(0)}(\mathbf{r},t)|^2
 \right)^2   \nonumber
\end{equation}
\normalsize Expanding the polynomial $P^{(1)}_l(\mathbf{r},t)$
around the central position $\mathbf{r}_{c  t}$, a parity argument
shows that the linear terms vanish:
\begin{eqnarray}
\label{eq:expression P} P^{(1)}_l(\mathbf{r},t) & = &   g_{I}
\left[ c_0(t) - c_{x}(t) (x-x_{c t})^2 \right. \nonumber \\
& \: & \left. \quad -c_{y}(t) (y-y_{c t})^2 - c_{z}(t) (z-z_{c
t})^2 \right] \nonumber
\end{eqnarray}
By definition of the lensing polynomial, the error function
\eqref{eq:lensing polynomial error function} must be stationary
with respect to each coefficient $c_{x,y,z,0}(t)\:$, which leads
to:
\begin{eqnarray}
\label{eq:coeffs polynomial} c_0(t)  =   \frac {7} {4 V(t)}, \: \:
c_{x,y,z}(t)  =  \frac {1} {2 w_{x,y,z \: t}^2 V(t)}, \nonumber \\
\mbox{with} \quad  V(t) = (2\pi)^{3/2} w_{x t} w_{y t} w_{z t}
\end{eqnarray}
Only the quadratic term intervene in the ABCD matrix: the
coefficient $c_0(t)$ merely adds a global additional phase to the
wave-function, which does not change the subsequent stability
analysis.

\subsection{Formal expression of the effective ABCD matrix.}

We can readily express the ABCD matrix associated with the evolution
under $\hat{H}^{(1)}(t)$. Writing this Hamiltonian in the form of
Eq.~\eqref{eq:introduction hamiltonien quadratique general}, and
using the formal expression~\eqref{eq:expression formelle matrice
ABCD} of the ABCD matrix as a time-ordered series, one obtains:
\begin{equation}
\label{eq:formal expression M1} M^{(1)}(t,t_0,X_0)= T \left[ \exp
\left(  \int_{t_0}^{t} dt' \left(
\begin{array}{cc}
\alpha(t') & \beta(t')  \\
\gamma(t') & \delta(t')  \\
\end{array} \right) \right) \right]
\end{equation}
In contrast to the usual linear ABCD matrices, this matrix now
depends on the input vector through the initial position width
matrix $X_0$  \footnotemark[3] \footnotetext[3]{The Hamiltonian
$\hat{H}^{(1)}(t)$ and the lensing polynomial
$P^{(1)}_l(\mathbf{r},t)$ depend of course also on $X_0$, but we do
not mention this dependence explicitly to alleviate the notations.}.
A brief inspection of Eq. \eqref{eq:introduction hamiltonien
quadratique general} and of the Hamiltonian $\hat{H}^{(1)}(t)$
\begin{eqnarray}
\hat{H}^{(1)}(t) & = & \frac {\hat{\mathbf{p}}^2} {2m} + m g
\hat{z}+ g_{I} \left[ c_0(t) \right. - c_{x}(t) (\hat{x}-x_{c
t})^2  \nonumber \\  & \: & \left. -c_{y}(t) (\hat{y}-y_{c t})^2 -
c_{z}(t) (\hat{z}-z_{c t})^2 \right],
\end{eqnarray}
shows that the matrices in the exponential read
$\alpha(t)=\delta(t)=0$, $\beta(t)=1$ and $\gamma(t)=\frac {2 g_{I}}
{m} \mbox{D}(c_x(t),c_y(t),c_z(t))$. Using Eq.~\eqref{eq:coeffs
polynomial}, one readily obtains the elements of the quadratic
matrix $\gamma$:
\begin{equation}
\gamma_{ii}(t)=  \frac {g_{I}} {m}  \frac {1} {w_{i t}^2
  (w_{x t} w_{y t} w_{z t})}
\end{equation}
for $i=x,y,z$, with the widths $w_{x,y,z t}$ given by
Eq.~\eqref{eq:linear width evolution}. A significant simplification
arises because $\gamma(t)$ is diagonal: one needs only to compute
the exponential of three $2 \times 2$ matrices associated with the
orthogonal directions $O_{x},O_{y},O_{z}$. The ABCD matrix is simply
the tensor product of those:
\begin{equation}
\label{eq:expression matrice ABCD H1} M^{(1)}(t,t_0,X_0) =
\otimes_{i=x,y,z} T \left[ \exp \left(  \int_{t_0}^{t} dt'
 \left(
\begin{array}{cc}
0 & 1  \\
\gamma_{ii}(t)
& 0  \\
\end{array}
 \right)   \right) \right]
\end{equation}

\subsection{Propagation in a a series of infinitesimal lenses.}

An infinitesimal expansion of~\eqref{eq:expression matrice ABCD H1}
shows that the evolution between
 $t$ and $t+dt$ is described by the ABCD matrix:
\begin{equation}
\label{eq:expression matrice ABCD H1} M^{(1)}(t+dt,t,X_0) \simeq
\left(
\begin{array}{cc}
1 & dt  \\
\gamma(t)\: dt
& 1  \\
\end{array} \right)
 \end{equation}
It can be rewritten as a product of two ABCD matrices:
\begin{equation}
\label{eq:expression matrice ABCD H1}
M^{(1)}(t+dt,t,X)
 \simeq
 \left(
\begin{array}{cc}
1 & dt  \\
0 & 1  \\
\end{array}
 \right)   \:
 \left(
\begin{array}{cc}
1 & 0  \\
\gamma(t) \: dt & 1  \\
\end{array}
 \right)
 \end{equation}
 If these were $2 \times 2$ matrices, in the optical formalism, the first matrix would be associated with the propagation of a ray
 on the length $dt$ and the second matrix, of the form
 \begin{equation}
\left(
\begin{array}{cc}
1 & 0  \\
-dt /f & 1  \\
\end{array}
 \right)
 \end{equation}
would model a lens of infinitesimal curvature $dt/f$. One can thus
consider, by analogy, that this second $6 \times 6$ matrix realizes
an atom-optical lens which curvature is the infinitesimal $3 \times
3$ matrix $\mbox{D}(\gamma_{xx}(t),\gamma_{yy}(t),\gamma_{zz}(t))\:
dt$. Besides, one can exploit the fact that it is a tensor product:
if one considers each direction $O_{x},O_{y},O_{z}$ separately, the
propagation amounts - as in optics - to a product of $2 \times 2$
matrices, which makes the analogy with a lens even more transparent.
The resulting $6 \times 6$  ABCD matrix is simply given by the
tensor product of those. Transverse degrees of freedom are,
nonetheless, coupled to each other through the lensing potential. It
is worth noticing that the focal lengths $f_{x},f_{y},f_{z}$ have
here the dimension of a time, and are negative if one considers
repulsive interactions: the quadratic potential
$P^{(1)}_l(\mathbf{r},\mathbf{p},t)$ acts as  a series of diverging
lenses associated with each infinitesimal time slice.

\subsection{Expression of the nonlinear ABCD matrix with the Magnus Expansion.}

Because of the time-dependence of the Hamiltonian
$\hat{H}^{(1)}(t)$,  the time-ordered exponential in \eqref{eq:expression matrice ABCD H1}
 cannot, in general, be
expressed analytically. Fortunately, a useful expression is
provided by the Magnus expansion~\cite{Magnus54}:
\begin{eqnarray}
\label{eq:expression martix M Magnus expansion} M^{(1)}(t,t_0,X_0)
& = &  \otimes_{i=x,y,z} \exp \left[ \int_{t_0}^{t} dt_1 N_i(t_1)
\right.
\nonumber \\
& \: & \left. + \frac {1} {2} {\int_0}^{t} dt_1 {\int_0}^{t_1}
dt_2 [ N_i(t_1),N_i(t_2)]+ ... \right] \nonumber \\ & \: & \:
\mbox{with} \: \: N_i(t)= \left(
\begin{array}{cc}0 & 1  \\
\gamma_{ii}(t) & 0 \\
\end{array} \right)
\end{eqnarray}
where $\otimes_{i=x,y,z}$ denotes again a tensor product. This
expansion has the advantage to preserve the
 unitarity of the evolution operator: at any order, the operator obtained by truncating the series
in the exponential is unitary. The Magnus expansion can be
considered as the continuous generalization of the Baker-Hausdorff
formula~\cite{Pechukas66} giving the exponential of a sum of two
operators $A$ and $B$ as a function of a series of commutators
along $\exp(A+B)= \exp A \exp B \exp ([A,B]/2)...$. The Magnus
expansion has been successfully applied to solve various physical
problems, among which differential equations in classical and
quantum mechanics~\cite{Marcus70}, spectral line
broadening~\cite{Cady74}, nuclear magnetic
resonance~\cite{Waugh82}, multiple photon
absorption~\cite{Schek81} and strong field effects in saturation
spectroscopy~\cite{BordePRA94}.\\

The first-order term $\Omega_1(t,t_0)$ in the argument of the
exponential can be expressed as
\begin{equation}
\label{eq: terms omega Magnus expansion} \Omega_1(t,t_0)  =
\int_{t_0}^{t} dt' N(t')= \left(
\begin{array}{cc}
0 & \tau  \\
 \langle \gamma \rangle \tau
& 0  \\
\end{array} \right) \quad \quad  \quad
\end{equation}
with the duration $\tau=t-t_0$ and the average quadratic diagonal
matrix $\langle\gamma \rangle_{ii}= 1/\tau \int_{t_0}^t dt
\gamma_{ii} (t)$. Exact expressions for $\langle\gamma
\rangle_{ii}$ are given in Eq.~\eqref{eq:calcul gamma cigar
condensate} of Appendix~\ref{app:cigar
 shaped} for a cylindrical
 condensate. Without this symmetry, the matrix elements $\langle\gamma
\rangle_{ii}$  cannot be evaluated
 analytically to our knowledge, but are nonetheless accessible with efficient
 numerical methods  \footnotemark[4] \footnotetext[4]{In the
short expansion limit considered later where $|t-t_0| \ll m
w^2_i(t_0)/ \hbar$, the average quantities $\langle \gamma_{ii}
\rangle$ can be approximated by the instantaneous value of the
quadratic coefficient $\gamma_{ii}$ at the center of the considered
time interval.}. The first-order ABCD matrix $M_1^{(1)}(t,t_0,X_0) =
e^{\Omega_1(t,t_0)} $ reads \footnotemark[5] \footnotetext[5]{For
repulsive interactions, all the eigenvalues of the matrix $\gamma$
are positive, and by convention its square root has also positive
eigenvalues.}:
\begin{equation}
\label{eq:first order ABCD matrix} M_1^{(1)} = \left(
\begin{array}{cc}
 \cosh (\langle\gamma \rangle^{1/2} \tau)  &  \langle \gamma \rangle^{-1/2}  \sinh
 (\langle \gamma \rangle^{1/2}  \tau )  \\ {\langle\gamma \rangle}^{1/2} \sinh (\langle \gamma \rangle^{1/2}  \tau
 )
  & \cosh ( \langle\gamma \rangle^{1/2} \tau )   \\
\end{array} \right)
\end{equation}
As expected, this main contribution of the Magnus expansion is
independent of the ordering of the successive infinitesimal lenses,
and can be interpreted as the ABCD matrix of a thick lens with
finite curvature. This expression is similar
 to the paraxial ABCD matrix  obtained in~\cite{LeCoq01} to describe the interactions between an atom laser
 and a condensate of known wave-function.\\

In the following developments, we use mainly this first-order
contribution to the Magnus
 expansion. In order to justify this approximation, we have performed a second-order computation of the ABCD matrix $M^{(1)}(t,t_0,X_0)$ in Appendix~\ref{app:second
 order}. This second-order correction
is weighted by the small parameter $\epsilon= (\tau/\tau_c)^{4}$,
depending on the ratio of the duration $\tau=t-t_0$ to a
time-scale $\tau_c$, which reads for a spherical cloud of radius
$w_0$:
\begin{equation}
\label{eq:time scale first order Magnus} \tau \ll \tau_c= \left(
\frac {w_0} {4 \pi a} \right)^{1/6} \frac {m w_0^{2}} { \hbar}
\end{equation}
One checks that the first-order expansion is valid for an arbitrary
long time ($\tau_c \rightarrow \infty$) as interaction effects
vanish ($a \rightarrow 0$). Considering a sample of initial radius
$w_0= 10 \: \mu \mbox{m}$, and using the s-wave scattering length $a
\simeq 5.7 nm$ of the $\:^{87}$Rb~\cite{LeCoq01}, one obtains
$\tau_c=0.31 \: s$. The convergence of the Magnus series is indeed
guaranteed when the following inequality is
satisfied~\cite{Pechukas66}:
\begin{equation}
N_m= \int_{t_0}^{t} dt'  \left\| N(t') \right\| < \ln (2),
\end{equation}
and our second-order computation gives an additional heuristic
indication of convergence for a flight duration $\tau \ll \tau_c$.

\section{STABILITY ANALYSIS OF A MATTER-WAVE RESONATOR}

In this Section, we apply the method of the ABCD matrix to discuss
the propagation of an atomic sample with mean-field repulsive
interactions in a matter-wave resonator~\cite{Impens06}. The
considered resonator involves a series of focusing atomic mirrors.
In this system, there is a competition between the transverse sample
confinement provided by the mirrors and the expansion induced by the
repulsive interactions, which determines the maximum size of the
sample during its propagation. In order to keep the sample within
the resonator, its transverse size must stay smaller than the
diameter of the laser beams realizing the atomic mirrors. If this
criterium is met during the successive bounces, the resonator is
considered as stable. The ABCD matrix method
 developed previously, giving an easy derivation of the sample
 width evolution, is well-suited to discuss this issue. One assumes an initial
 Gaussian profile for the sample wave-function. The atomic wave propagation
 in-between the mirrors is treated in the aberrationless
 approximation,  and described by the nonlinear ABCD matrix~\eqref{eq:first order ABCD matrix} accounting for self-interaction effects. The
evolution of the sample width obtained with this method is compared
to the behavior expected from a non-perturbative paraxial approach.


\subsection{Resonator description}

The considered matter-wave resonator is based on the levitation of
a free-falling two-level atomic sample by periodic vertical Raman
light pulses. This proposal is described in detail in the
reference~\cite{Impens06}, but we remind here its main features
for the sake of clarity. In the absence of light field, the atomic
sample propagates in the Hamiltonian~\eqref{eq:basic Hamiltonian}.
We consider an elementary sequence which consists in a pair of two
successive short vertical Raman $\pi$ pulses~\cite{BermanBook}.
Each pulse is performed by two counter-propagating laser beams of
respective frequencies $\omega_{up}$, $\omega_{down}$ and
wave-vectors $k_{up}=k \mathbf{z}$, $k_{down}= - k \mathbf{z}$
equal in norm to a very good approximation and of opposite
orientation. The first Raman pulse propagates upward with an
effective vertical wave-vector $\mathbf{k}_{e,1}=2  k \mathbf{z}$
and corresponds to laser frequencies $\omega_{up}=\omega_2$, \:
$\omega_{down}=\omega_1$; the second one propagates downward with
an effective vertical wave-vector $\mathbf{k}_{e,2}=-2  k
\mathbf{z}$ and with the laser frequencies $\omega_{up}=\omega_3$
and $\omega_{down}=\omega_4$. The frequencies $\omega_{1,2,3,4}$
are adjusted so that both Raman pulses have the same effective
frequency $\omega_{e}=|\omega_{up}-\omega_{down}|$, satisfying the
resonance condition~\cite{Impens06} $\omega_e=\omega_2 - \omega_1
= \omega_4 - \omega_3 = \omega_{ba}- 2 \hbar k^2 /(m \hbar)$. The
intermediate level involved during the Raman pulses (of energy
$E=\hbar (\omega_a+ \omega_e)$) is taken sufficiently far-detuned
from the other atomic energy levels to make spontaneous emission
negligible~\footnotemark[6] \footnotetext[6]{In practice, a
detuning on the order of the GHz - experimentally compatible with
$\pi$ pulse of duration shorter than the ms~\cite{Gauguet08} - is
sufficient to discard spontaneous emission.}. After adiabatic
elimination of the intermediate level, the action of the Raman
pulses can be modelled by the effective dipolar Hamiltonian:
\begin{equation}
\label{eq:dipolar potential accelerated frame} \hat{H}_{dip}(t) =
- \hbar \Omega_{ba}(\mathbf{r},t ) \: \cos ( \omega_e t -
\mathbf{k}_{e1,2} \cdot \hat{\mathbf{r}}  ) \: ( | b \rangle
\langle a | + | a \rangle \langle b | )
\end{equation}
Each pair of pulses acts as an atomic mirror, bringing back the
atoms in their initial internal state $a$, and providing them with
a net momentum transfer of $\Delta \mathbf{p} \simeq 4 \hbar
\mathbf{k}$. The atomic motion is sketched on Fig.~\ref{fig:phase
space} in the energy-momentum picture.
\begin{figure}[htbp]
\begin{center}
\includegraphics[width=6cm]{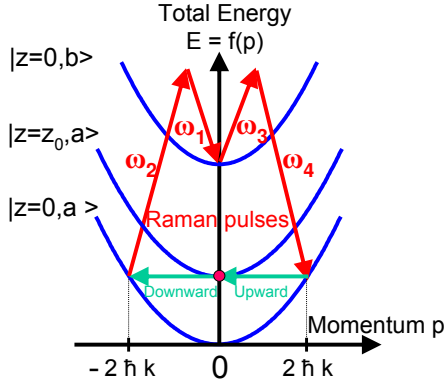}
\end{center}
\caption{(Color
  online)Evolution of the atomic sample in the energy-momentum
picture. The total energy includes the kinetic, gravitational and
internal energy. The atoms are initially at rest $(p=0)$, at the
altitude $z_0$, and in the lower state $a$. The starting point is
thus at the intersection of the paraboloid $(a,z_0)$ and of the
vertical axis (p=0). In between the pulses, the motion of the
atomic sample in the gravity field is conservative: it corresponds
to a leftward horizontal trajectory of the representative point.}
\label{fig:phase space}
\end{figure}\\
 This sequence can be repeated many
times. If the period $T$ in-between two successive atomic mirrors
is set to
\begin{equation}
\label{eq:resonant period} T := T_0 = \frac  {4 \hbar k} {m g} \:,
\end{equation}
the acceleration provided by the Raman pulses compensates on average
that of gravity: the cloud levitates and evolves inside a
matter-wave resonator~\cite{Impens06}. An analogous system has been
realized experimentally recently~\cite{Hughes09}.

\subsection{Focusing with atomic mirrors.}

Matter-wave focusing can be obtained, in principle, with laser
waves of quadratic intensity profile~\cite{Whyte04,Murray05} or
alternatively of spherical wave-front~\cite{Impens06}. We
concentrate on the focusing obtained with an electric field of
quadratic intensity profile~\cite{Murray05}, the discussion of
which is less technical. The Rabi frequency considered for the
Raman pulses of the resonator depends quadratically on the
distance to the propagation axis $O_z$~\footnotemark[7]
\footnotetext[7]{Close to the propagation axis, this quadratic
profile can be reproduced to a good approximation with Raman
pulses of Gaussian intensity profile.}:
\begin{equation}
\label{eq:rabi quadratique} \Omega_{ba}(x,y,z,t )=  \left(1- \frac
{x^2+y^2} {2 w_{las}^2} \right) \Omega_0(t)
\end{equation}
These Raman pulses generate a quadratic position-dependent
light-shift proportional to the field intensity and thus to the
square of the Rabi frequency~\eqref{eq:rabi quadratique}. After
the pulse, the atomic wave-function initially in of the form of
Eq.~\eqref{eq:fonction initiale gaussienne} is thus multiplied by
a factor yielding the input-output relation:
\begin{equation}
\label{eq:reflection input output} \psi_{out}(\mathbf{r},t)= e^{i 2
k (z-z_0)} e^{- i (x^2+y^2) / w_{las}^2} e^{i \phi'_{0}}
\psi_{in}(\mathbf{r},t)
\end{equation}
with $\phi'_{0}$ a constant phase added at the condensate center
$\mathbf{r}_0=(0,0,z_0)$ during the pulse. The outgoing
wave-function can thus be put again in the form of
Eq.~\eqref{eq:fonction initiale gaussienne} if one replaces $p_0$ by
$p_1=p_0+ 2 \hbar k$, and $X_0, Y_0$ with
\begin{equation}
\label{eq:ABCD transfo mirror} \left(
\begin{array}{c}
X_1 \\
Y_1
\end{array} \right)=
 \left( \begin{array}{cc}
I_{3} & 0_{3}  \\
   D(-1 / f,-1/f,0)
& I_{3}  \\
\end{array}
\right) \left(
\begin{array}{c}
X_0 \\
Y_0
\end{array} \right)
\end{equation}
$I_{3},0_{3}$ are the $3\times3$ identity matrix and null matrix,
$D(-1 / f,-1/f,0)$ is as previously a $3\times3$ diagonal matrix.
The focal time is:
\begin{equation}
\label{eq:distance focale}
 f \: = \: \frac {m w_{las}^2} {2 \: \hbar }
\end{equation}
Eq.~\eqref{eq:ABCD transfo mirror} shows that the pulse acts as a
lens in the transverse directions $O_x,O_y$ \footnotemark[8]
\footnotetext[8]{The absence of focusing in the direction of laser
beam propagation $O_z$ is not critical since it does not drive the
cloud out of the beam.}. \\

The strength of the focusing which can be achieved with such
atomic mirrors~\footnotemark[9] \footnotetext[9]{The considered
atomic mirrors consist indeed not in a single, but in a double
Raman pulse. This does not change the qualitative discussion of
this paragraph.} is indeed limited by the quasi-uniformity
required for the Rabi frequency on the condensate surface, in
order to perform an efficient population transfer with the Raman
$\pi$-pulse. Considering a cigar-shaped cloud of small width
$w_{\bot}$ along the $O_x,O_y$ axis, one may require that the Rabi
frequency difference between the border and the center of the
cloud satisfies: $|\Omega(w_{\bot},0,z,t)-\Omega_0(t)| /
|\Omega_0(t)| \leq \epsilon$. This yields readily a lower bound on
the focal time $f$:
\begin{equation}
 f \: \geq \: \frac {m \:w^2_{\bot}} {2 \: \hbar \: \epsilon}
\end{equation}
With a reasonable bound of $\epsilon=10^{-2}$, a cylindrical cloud
of $\:^{87}$Rb atoms of transverse size $w_{\bot} \simeq \: 10 \:
\mu m$, one obtains a minimum focusing time: $f \: \geq \: 6.7 s$. A
back-on-the-envelope computation of the reflection coefficient shows
that the losses resulting from such an inhomogeneity of the Rabi
frequency are on the order of $10^{-3}$.

\subsection{Resonator stability analysis.}

We now investigate the non-linear ABCD propagation of a cigar-shaped
sample in the resonator.  As a specific example, we consider a cloud
of $^{87}$Rb atoms  taken in the two internal levels $|a \rangle =
|5 S_{1/2}, F=1 \rangle $ and $|b \rangle =| 5 S_{1/2}, F=2
\rangle$. In-between the Raman mirrors, the whole sample is expected
to propagate in the ground state $|a \rangle$. We consider a sample
of $N=10^5$ atoms, of initial dimensions $w_{x}=w_{y}=w_{r}= 10 \:
\mu m$ and $w_{z}=100 \: \mu m $, and we use the s-wave scattering
length $a \simeq 5.7 \: nm$ of the Rubidium. We investigate the
evolution of this sample during a thousand bounces and for various
mirror focal times. Keeping a significant atomic population inside a
matter-wave resonator during such a big number of reflections is
challenging, but not impossible in principle given the high
population transfer which has been achieved experimentally with
Raman pulses~\cite{WeitzPRA94}~\footnotemark[10]
\footnotetext[10]{We treat the wave-propagation in the resonator as
if the atomic cloud was entirely reflected on the successive atomic
mirrors. Indeed, even if resonant Raman pulses can perform a
population transfer with an efficiency close to
99\%~\cite{WeitzPRA94}, the residual losses become significant after
a big number of bounces in a real experiment. This results in a
gradual decrease of the mean-field interactions, which could be
accounted for in a more sophisticated model. Our point here is
simply to illustrate the nonlinear ABCD method on a thought
experiment, and we thus adopted a simplified approach with perfect
atomic mirrors.} One obtains the value $T_0 \simeq 1.5 \: ms$ for
the period between the Raman mirrors. This time scale is much
shorter than the duration $\tau_c \simeq 0.3 \: s$ found for the
validity of the first-order Magnus expansion associated with a
spherical cloud of radius $w_0= 10 \: \mu \mbox{m}$. This shows that
the ABCD matrix of the cigar-shaped condensate is well-approximated
by the leading order [Eq.~\eqref{eq:first order ABCD matrix}] of the
Magnus expansion~\footnotemark[11] \footnotetext[11]{We have
computed the time-scale $\tau_c$ determining the validity of the
first-order Magnus term for spherical wave-packets only.
Nonetheless, a basic dimensional analysis shows that for a
cigar-shaped cloud, the time-scale determining the validity of the
first-order Magnus term is bounded below by the time $\tau_c$ given
by Eq.~\eqref{eq:time scale first order Magnus} and computed by
setting $w_0$ equal to the smallest cigar dimension.}. Furthermore,
the free-propagation time $T_0$ is also much shorter than the
time-scale $\tau_r= m w^2_{r}/ \hbar$ associated with the free
expansion of the transverse width, so that one can safely
approximate the average quadratic coefficient $\langle\gamma
\rangle$ with the instantaneous value $\langle\gamma \rangle \simeq
\gamma(w_{x \: T_0/2},w_{y \: T_0/2},w_{z \: T_0/2})$.\\

\begin{figure}[htbp]
\begin{center}
\includegraphics[width=6.5cm]{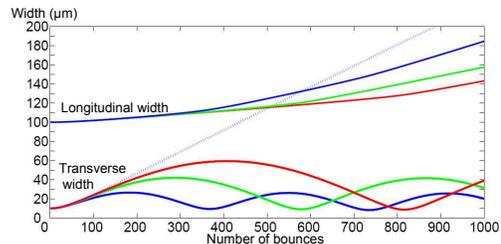}
\end{center} \caption{(Color
  online)Evolution of the transverse and longitudinal
width of the sample ($\mu m$) during the successive bounces in the
cavity (numbered from 1 to 1000), for the mirror focal times $f=20
\: s$ (blue), $f=50 \: s$ (green) and $f=100 \: s$ (red). The
dashed line represents the evolution of the transverse width in
the absence of focusing with the Raman mirrors.}
\label{fig:evolution condensat}
\end{figure}

To compute the evolution of the transverse and longitudinal sample
width, one proceeds as follows. As in
Section~\ref{subsec:determination lensing pragraph}, one starts with
initial width matrices $X_0= i \mbox{D}( w_{x 0}, w_{y 0}, w_{z 0})$
and $Y_0= \frac {\hbar} {m} \mbox{D} (1/w_{x 0}, 1 / w_{y 0}, 1 /
w_{z 0})$ and computes the interacting ABCD matrix~\eqref{eq:first
order ABCD matrix} as a function of these initial widths. During the
first cycle, one multiplies the corresponding vector $(X_0,Y_0)$
successively with nonlinear ABCD matrix~\eqref{eq:first order ABCD
matrix} and with the mirror ABCD matrix~\eqref{eq:ABCD transfo
mirror}. The new width matrices $(X_1,Y_1)$ are obtained, from which
one can infer the nonlinear ABCD matrix for the next propagation
stage. The iteration of these algebraic operations is a
straightforward numerical task. The results, depicted on
Fig.~\ref{fig:evolution condensat}, show that the transverse width
oscillates with an amplitude and a period which both increase with
the mirror focal time. The maximum sample size is $w_r= 25 \: \mu m$
and $w_r=  60 \: \mu m$ for the respective focal times $f= 20 \: s$
and $f= 100 \:s$. Considering for instance a laser beam of waist
$w=100 \: \mu m$ in the experiment, one sees that with those focal
times the atomic cloud remains within the light beam and is thus
efficiently confined transversally in the resonator. As expected,
the use of Raman mirrors with a stronger curvature allows one to
shrink the transverse size of the stabilized cloud.
Fig.~\ref{fig:relation focale taille} shows the evolution of the
maximum sample transverse size as a function of the mirror focal
time. The extended ABCD matrix analysis presented in this paper
allows thus to determine efficiently the minimum amount of focusing
required to keep the sample within the diameter of the considered
Raman lasers. In that respect it can be used to optimize the
trade-off, exposed in the previous paragraph,
between strongly focusing or highly reflecting atomic mirrors.\\

\begin{figure}[htbp]
\begin{center}
\includegraphics[width=6.5cm]{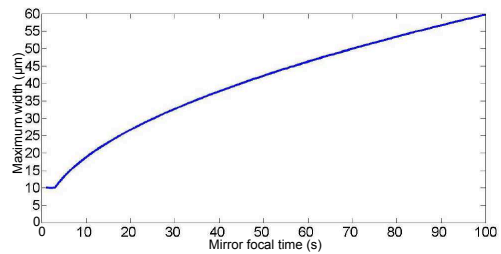}
\end{center} \caption{(Color
  online)Maximum sample transverse width ($\mu m$)
during the evolution in the resonator as a function of the Raman
mirror focal time ($s$). We have considered the first $1000$
bounces to determine this maximum.} \label{fig:relation focale
taille}
\end{figure}

\subsection{Comparison with the predictions of the nonlinear paraxial equation.}
\label{subsec:comparison aberationless paraxial}

As exposed in Appendix~\ref{app:belanger}, the propagation of an
atomic beam with a longitudinal momentum much greater than the
transverse momenta can be alternatively described by a paraxial
wave equation of the form~\eqref{eq:eqparaxial}. Furthermore, if
the linear density of the atomic beam is uniform, the nonlinear
coefficient intervening in this paraxial equation is a constant.
As in nonlinear optics~\cite{Pare92} and in 2D
condensates~\cite{Pitaevskii97}, this equation induces a universal
behavior in paraxial atomic beams~\cite{ImpensGuidedAtomLaser08}:
the transverse width oscillates with a frequency independent from
the strength of the interaction. The width oscillations, depicted
on Fig.~\ref{fig:evolution condensat}, indeed allow one to
confront the results of our method, which uses a non-paraxial wave
equation treated in the aberrationless approximation, to the
predictions of the full nonlinear paraxial equation with a uniform
nonlinear coefficient. We stress that this second approach leaves
the nonlinear term as such and does not assume that the Gaussian
shape of the atomic beam is preserved. In this sense it is more
exact than the radius of curvature method used in
Appendix~\ref{app:belanger}. It is also approximate, since the
atomic beam is neither paraxial nor of uniform linear density.
Nevertheless, it is remarkable that
both treatments agree very well on the oscillation period of the width.\\

To apply the paraxial description, one models the action of the
successive mirrors on the transverse wave-function with an average
potential. The lens operated by each Raman mirror, of focal time
$f$, imprints a phase factor of $e^{i \frac {m} {2 \hbar f} r^2}$
[see Eq.~\eqref{eq:reflection input output} and
Eq.~\eqref{eq:distance focale}]. The series of lenses, separated by
the duration $T_0$, thus mimics the following effective quadratic
potential:
\begin{equation}
V_{\bot, lens}=  \frac {m} {2 \hbar^2 T_0 f}  r^2
\end{equation}
Let us consider the nonlinear contribution, given by a contact
term of the form $V_{\bot, int}(\mathbf{r})= g_I
|\psi_{\!/\!/}(z)|^2 |\psi_{\perp}(\mathbf{r})|^2
\psi_{\perp}(\mathbf{r})$, with $\psi_{\!/\!/}(z)$ the
longitudinal wave-function [Eq.~\eqref{eq:psiparr} of
Appendix~\ref{app:belanger}]. The term $g_I |\psi_{\!/\!/}(z)|^2$
appears as an effective nonlinear coupling coefficient for the
transverse wave-function depending on the altitude $z$. Adding
this nonlinear contribution to Eq.~\eqref{eq:eqparaxial0}, one
obtains a 2D nonlinear Schr\"odinger equation (NLSE):
\begin{eqnarray}
    i \hbar\partial_\zeta \psi_{\perp}(x,y,\zeta) & = & \left[- \frac {\hbar^2}{2m}(\partial_x^2+\partial_y^2)
    + g_I |\psi_{\!/\!/}(\zeta)|^2 |\psi_{\perp}|^2 \right. \nonumber \\
    &\: & \quad \left. + \frac {m} {2 \hbar^2 T_0 f}  r^2\right]
    \psi_{\perp}(x,y,\zeta)\,,
\end{eqnarray}
$\zeta$ is a parameter defined in Eq.~\eqref{eq:definition zeta}
equivalent to the propagation time, $a$ the scattering length and
$r^2=x^2+y^2$. Setting $K=\frac {m} {\hbar}$, one can recast this
equation in the same form as in~\cite{Pare92} where the propagation
of a light wave in a quadratic graded index medium was considered:
\begin{equation}
2  i   K   \partial_\zeta \psi_{\perp} = \left[ -
\partial_T^2+ \left( 8 \pi a |\psi_{\!/\!/}|^2 \right) |\psi_{\perp}|^2 +K^2 \left( \frac {1} {f T_0} \right) r^2 \right] \psi_{\perp} \,,
\label{eq:paraxial propagation comparaison}
\end{equation}
We now make the assumption that the variations of the non-linear
coefficient $8 \pi a |\psi_{\!/\!/}(\zeta)|^2$ with $\zeta$ are
sufficiently smooth to have a negligible impact on the period of
the sample width oscillations. This assumption seems reasonable
for the considered cigar-shaped cloud, which has a slow
longitudinal expansion in comparison with the oscillation period
[see Fig.~\ref{fig:evolution condensat}]. This hypothesis is
indeed validated \textit{a posteriori}, since it leads to
predictions in excellent agreement with the results of the ABCD
method discussed above. Once the nonlinear coefficient is
approximated with a constant, one can readily apply the results
derived in~\cite{Pare92,ImpensGuidedAtomLaser08}, which show that
Eq.~\eqref{eq:paraxial propagation comparaison} yields transverse
oscillations of universal frequency:
\begin{equation}
\omega_{par}= \frac {2} {\sqrt{f T_0}} \label{eq:universal
frequency}
\end{equation}
The results obtained from the perturbative ABCD approach are
confronted with this prediction on Fig.~\ref{fig:comparaison periode
oscillation}. The agreement improves as the mirror focal time
increases, and it is in fact already good ($4\%$) for a focal time
of $f = 3 \: s$ and attains $0.7\%$ for a focal time of $f=50 \: s$.
As discussed above, focal times shorter than $f=20 \: s$ seem
incompatible with the reflection coefficient desired for the atomic
mirrors. The disagreement observed below $f \leq 3 \: s$ may be
attributed to a failure of the paraxial approximation to describe
the propagation of the sample in our system.
\begin{figure}[htbp]
\begin{center}
\includegraphics[width=6.5cm]{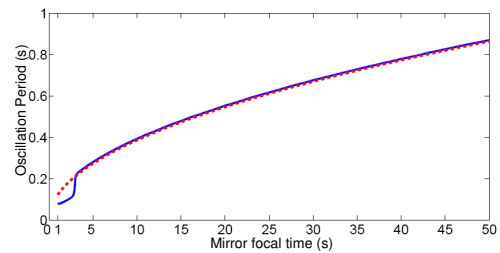}
\end{center} \caption{(Color
  online)Period of the transverse width oscillations ($s$)
 in the matter-wave resonator as a function of the Raman
mirror focal time ($s$). The full and the dashed line give the
oscillation periods obtained respectively through the perturbative
ABCD approach and through the nonlinear paraxial wave equation.}
\label{fig:comparaison periode oscillation}
\end{figure}


\section{CONCLUSION}

This paper exposed a treatment of the non-linear Schr\"odinger
equation involving theoretical tools from optics and atom-optics.
The ABCD propagation method for matter waves has been extended
beyond the linear regime thanks to a perturbative analysis relying
on an atom-optical aberrationless approximation. We have derived
approximate analytical expressions for the ABCD matrix of an
interacting atomic cloud thanks to a Magnus expansion. This matrix
analysis has been applied to discuss the propagation of an atomic
sample in a perfect matter-wave resonator. We have shown that such
sample can be efficiently stabilized thanks to focusing atomic
mirrors. We have found that the nonlinear ABCD propagation
reproduces to a good level of accuracy the universal oscillations
expected from the nonlinear paraxial equation for matter
waves~\cite{ImpensGuidedAtomLaser08}, which makes it a promising
tool to model future nonlinear atom optics experiments and a
seducing alternative to previous numerical methods applied to
matter-wave resonators~\cite{Whyte04}. We have also highlighted an
other optical method, involving more stringent assumptions -
paraxial propagation, cylindrical symmetry and constant longitudinal
velocity - and also relying on the aberrationless approximation.
This last method enables one to address self-interaction effects in
the free propagation through a complex parameter [defined in
Eq.~\eqref{eq:nonlinear q}], which is analogous to a radius of
curvature, and the evolution of which is very simple
[Eq.~\eqref{eq:evolution q nonlineaire}]. As far as the beam width
is concerned, the effect of self-interactions can be interpreted as
a scaling transformation of the free propagation by a factor
depending on the matter-wave flux $\mathcal{F}$ [See
Eq.~\eqref{eq:width expansion}]. Both approaches are relevant to
study interaction effects on the stability of atomic sensors resting
on Bloch oscillations~\cite{Biraben05}, on the sample propagation in
coherent interferometers~\cite{CoherentAtomInt}. An interesting
continuation of this work would be to develop a nonlinear ABCD
matrix analysis beyond the aberrationless approximation.

\section*{ACKNOWLEDGEMENTS}
The authors acknowledge enlightening discussions with Yann Le Coq
on the nonlinear paraxial equation for matter waves. Fran\c{c}ois
Impens thanks Nicim Zagury and Luiz Davidovich for their
hospitality. This work was supported by DGA (Contract No 0860003)
and by CNRS. Our research teams in SYRTE and Laboratoire de
Physique des Lasers are members of IFRAF(www.ifraf.org).\\

\appendix

\section{The Method of the Non-Linear Radius of Curvature.}
\label{app:belanger}

This method adresses the paraxial propagation of a monochromatic
and cylindrical matter-wave beam. It relies on the introduction of
an effective complex radius of
curvature~\cite{Kogelnik65,Yariv78}, which evolution is especially
simple, even for a self-interacting beam. It has been applied
 successfully by B\'elanger and Par\'e~\cite{Belanger83} to describe self focusing
phenomena of cylindrical optical beams propagating in the paraxial
approximation, and it works equally well for matter waves
propagating in the same regime. This is typically the case for an
atom laser beam falling into the gravity field, for which the
transverse momentum components become negligible
compared to the vertical momentum after sufficient time~\cite{LeCoq01}.\\

We consider a mono-energetic wave-packet propagating in the
paraxial regime, and evolving in the sum of a longitudinal
potential $V_{/\!/}(z)$ and a transverse one $V_{\perp}(x,y,z)$,
which may also vary slowly with the longitudinal coordinate $z$.
This section begins with a brief remainder on the paraxial
equation for matter waves~\cite{Riou08}, and on its spherical-wave
solutions in the linear case~\cite{Yariv78}. It is remarkable that
such solutions can be extended to the nonlinear
propagation~\cite{Yariv78}, at the cost of certain approximations,
and thanks to the introduction of a generalized radius of
curvature depending on the coupling strength. Our treatement of
the nonlinear matter wave propagation follows step by step the
approach of B\'elanger and Par\'e for optical
waves~\cite{Belanger83}.

\subsection{The Paraxial Equation for Matter Waves.}

Our derivation of the nonlinear paraxial wave-equation follows the
treatment done in~\cite{Riou08}. The wave-function is factorized
into a transverse and longitudinal component:
\begin{equation}
\psi(x,y,z)=\psi_{\perp}(x,y,z)\,\psi_{\!/\!/}(z) \nonumber
\end{equation}
The longitudinal component obeys a 1D time-independent
Schr\"odinger equation,
\begin{equation}
-\frac{\hbar^2}{2m}\frac{\partial^2 \psi_{\!/\!/}}{\partial z^2}
+V_{/\!/}\psi_{\!/\!/}=E\psi_{\!/\!/}\,, \label{eq:Schro1D}
\nonumber
\end{equation}
 which can be solved with the WKB method:
\begin{equation}
\psi_{\!/\!/}(z)= \sqrt{\frac {m \mathcal{F}} {p(z)}}
\exp\!\left[\frac{i}{\hbar}\int_{z_0}^z \mathrm{d}u\,p(u)
\right]\,. \label{eq:psiparr} \nonumber
\end{equation}
$\mathcal{F}=\int d^2 \mathbf{r}_{\bot} \frac {p(z)} {m} |\psi(
\mathbf{r}_{\bot},z)|^2 $ is the atomic flux evaluated through any
infinite transverse plane, the transverse wave-function
$\psi_{\perp}$ being normalized to unity $\int d^2
\mathbf{r}_{\bot} |\psi_{\perp}(\mathbf{r}_{\bot},z)|^2=1$.
$p(z)=\sqrt{2m\left(E - V_{/\!/}(z)\right)}$ is the classical
momentum along $z$, and $z_0$ is the associated classical turning
point verifying $p(z_0)=0$. The transverse wave-function
$\psi_{\perp}$, assumed to depend slowly enough on the coordinate
$z$ to make its second derivative negligible, verifies the
equation:
\begin{equation}
\left[i\hbar \frac {p(z)} {m} \partial_z + \frac
{\hbar^2}{2m}(\partial_x^2+\partial_y^2) - V_\perp(x,y,z)\right]
\psi_{\perp}(x,y,z)=0\, .
    \label{eq:eqparaxial0} \nonumber
\end{equation}
This equation can be simplified with a variable change in which
the longitudinal coordinate $z$ is replaced by the parameter
$\zeta$:
\begin{equation}
\label{eq:definition zeta}
 \zeta(z) = \int_{z_0}^{z}\!dz\, \frac {m} {p(z)}
\end{equation}
which corresponds to the time needed classically to propagate from
the turning point $z_0$ to the coordinate $z$ \footnotemark[12]
\footnotetext[12]{Indeed, this parameter appear as proportional to
the proper time experienced by the atom on the classical
trajectory determined by $p(z)$~\cite{BordeVarenna07}.}. The wave
equation becomes
\begin{equation}
    \left[i\hbar\partial_\zeta + \frac {\hbar^2}{2m}(\partial_x^2+\partial_y^2) - V_\perp(x,y,\zeta)\right] \psi_{\perp}(x,y,\zeta)=0\,,
    \label{eq:eqparaxial}
\end{equation}
We assume from now on that the transverse potential
$V_{\perp}(x,y,z)$ has a cylindrical symmetry. If one sets $K= m /
\hbar$ and $V_\perp(x,y,\zeta)= \frac {\hbar} {2} K_2(\zeta) r^2$
with $r^2=x^2+y^2$, Eq.~\eqref{eq:eqparaxial} has the same form as
the paraxial
 equation for the electric field used in~\cite{Belanger83}:
\begin{equation}
    \left[\partial_T^2 \: + \: 2  i   K   \partial_\zeta \: - \:
    K
K_2(\zeta) r^2 \right] \psi_{\perp}(x,y,\zeta)=0\,,
    \label{eq:eqparaxial2}
\end{equation}
It is worth noticing that, as a consequence of our variable
change, the derivative with respect to the longitudinal coordinate
$z$ has been replaced by a time derivative with respect to $\zeta$.\\

\subsection{Spherical Wave solutions to the linear equation.}

One looks for solutions of Eq.~\eqref{eq:eqparaxial2} of the kind:
\begin{equation}
\label{eq:spherical wave function} \psi_{\perp} ( x, y, \zeta ) =
A( \zeta ) \exp \left[ i \frac {K} {2 q(\zeta)} r^2 \right]
\end{equation}
with again $K=m/\hbar$. Such function is a solution if and only if
the parameter $q(\zeta)$ - called complex radius of curvature, and
homogenous to a time for matter waves - satisfies the following
equation:
\begin{equation}
\label{eq:equation on q} \frac {q'-1} {q^2} - \frac {K_2 ( \zeta
)} {K} = 0
\end{equation}
and if the amplitude $A(\zeta)$ verifies:
\begin{equation}
\label{eq:equation on A} \frac {A'} {A} + \frac {1} {q}  = 0
\end{equation}
The prime stands for the derivative with respect to $\zeta$. In
the absence of the transverse potential, i.e.
$V_\perp(x,y,\zeta)=0$, an obvious evolution is obtained with
 $q(\zeta)=\zeta$.\\

These equations imply a relation between the amplitude and width
of the wave-function. We adopt the usual decomposition for the
complex radius of curvature along its imaginary and complex part:
\begin{equation}
\frac {1} {q} = \frac {1} {R} + \frac {2 i } {K w^2} \nonumber
\end{equation}
 Assuming that $K_2(\zeta)$ is real, and
 combining the  imaginary
part of  Eq.~\eqref{eq:equation on q} with the real part of
Eq.~\eqref{eq:equation on A}, one obtains:
\begin{equation}
 |A(\zeta)|^2=  |A_0|^2 \frac {w_0^2} {w^2(\zeta)} \nonumber
\end{equation}
This relation reflects the conservation of the atomic flux
$\mathcal{F}$ along the propagation. With our choice of
normalization, the parameter $|A|^2$ is given by:
\begin{equation}
\label{eq:flux conservation} |A|^2= \frac {2} {\pi  w^2}
\end{equation}

\subsection{Spherical Wave solutions to the nonlinear equation.}

With several approximations, it is possible to find similar
solutions in the interacting case. Atomic interactions are
described by the mean-field potential
 \begin{equation}
 V_{i}(x,y,\zeta)= g_{I}^0  |\psi_{/\!/}(z)|^2
 |\psi_{\perp}(x,y,\zeta)|^2 \quad \mbox{with} \quad g_{I}^0= \frac {4 \pi \hbar^2 a} {m} \nonumber
 \end{equation}
which intervenes in the time-independent equation verified by
$\psi$. Because we adopt here a different normalisation for the
wave-function, the nonlinear coupling constant $g_{I}^0$ differs
from the coupling constant $g_I$ used previously: $g_{I}^0=g_I/N$.
The mean-field contribution induces the following transverse
potential
\begin{equation}
V_{\perp}(x,y,\zeta)= g_{I}^0  |\psi_{/\!/}(\zeta)|^2
\left(|\psi_{\perp}(x,y,\zeta)|^2- |\psi_{\perp}(0,0,\zeta)|^2
\right) \nonumber
\end{equation}
in the paraxial equation verified by  $\psi_{\bot}$. In the
considered example, this potential receives no other contribution.
The subsequent analysis requires three important approximations.
First, it uses the ``aberrationless approximation'', which assumes
that the wave-function follows the Gaussian profile
\eqref{eq:spherical wave function} in spite of the non-linearity.
Second, it assumes that the transverse mean-field potential is
well-described by a second order expansion,
\begin{equation}
V_{\perp}(x,y,\zeta) \simeq - 2 g_{I}^0  |\psi_{/\!/}(\zeta)|^2
\frac {|A(\zeta)|^2} {w^2(\zeta)} \: r^2 \label{eq:belanger
interacttion potential}
\end{equation}
The term $G(\zeta)=g_{I}^0 |\psi_{/\!/}(\zeta)|^2 $ can be seen as
the atom-optical equivalent of a third-order non-linear
permittivity. Third, it neglects the dependence on $G(\zeta)$
towards the altitude, which is a valid approach if the linear
density $n_{1D}=m\mathcal{F}/p(z)$ is a constant~\footnotemark[13]
\footnotetext[13]{This approximation is indeed implicit in the
treatment of B\'elanger and Par\'e~\cite{Belanger83}, since it is
necessary to obtain the nonlinear paraxial wave-equation which is
the starting point of their analysis.}. We assume from now on that
the atomic flux $\mathcal{F}$ is constant and that the average
longitudinal momentum $p(z)=\sqrt{2m\left(E - V_{/\!/}(z)\right)}
\simeq p_{0 /\!/ }$ varies very slowly with $z$. The parameter
$\zeta$ can then be expressed simply as $\zeta= m (z-z_0) / p_{0
/\!/ }$. Eq.~\eqref{eq:belanger interacttion potential} and the
normalization of $\psi_{\perp}$ [Eq.~\eqref{eq:flux conservation}]
give readily:
\begin{equation}
\label{eq:expression K2} \frac {K_2 ( \zeta )} {K}=  \frac {-8
g_{I}^0 \mathcal{F}} {\pi p_{0 /\!/ } w^4(\zeta)} \nonumber
\end{equation}
Eq.~\eqref{eq:equation on q} can then be recast as:
\begin{equation}
\label{eq:equation on q 2} \frac {q'-1} {q^2} + \frac
{\mathcal{F}} {\mathcal{F}_c} \left( \frac {4} {K^2 w^4(\zeta)}
\right) =0 \nonumber
\end{equation}
The quantity $\mathcal{F}_c$, called critical flux, reads
$\mathcal{F}_c= \pi p_{0 /\!/ } \hbar^2 / (2  g_{I}^0 m^2)$. The
last equation may be split into its real and imaginary part along:
\begin{equation}
\label{eq:equation on q 2 real} \left( \frac {1} {R} \right)' +
\frac {1} {R^2}-\sigma  \left( \frac {2} {K w^2} \right)^2 =0
\end{equation}
and
\begin{equation}
\label{eq:equation on q 2 imaginary}
 \left( \frac {1} {K w^2} \right)' + 2 \left( \frac {1} {R} \right) \left( \frac {2} {K w^2}
\right)  =0
\end{equation}
where we have introduced the dimensionless parameter $\sigma = 1 +
\mathcal{F}/ \mathcal{F}_c$. This system can be uncoupled thanks
to the following trick: Eq.~\eqref{eq:equation on q 2 imaginary}
is multiplied by $i \sqrt{\sigma}$ and added to
Eq.~\eqref{eq:equation on q 2 real}. One obtains:
\begin{equation}
\label{eq:equation on q 3}
 \left( \frac {1} {R} \right)'
 + i \sqrt{\sigma}  \left( \frac {2} {K w^2}
\right)'  + \frac {1} {R^2}+ \frac {2 i \sqrt{\sigma}} {R}  \left(
\frac {2} {K w^2} \right) - \sigma \left( \frac {2} {K w^2}
\right)^2 =0
\end{equation}
This equation can be simply interpreted as
\begin{equation}
 q'_{NL}-1 =0
\end{equation}
with the generalized complex radius of curvature:
\begin{equation}
\label{eq:nonlinear q} q_{NL} = \frac {1} {R} + \frac {
  2 \sqrt{\sigma} i} {K w^2}
\end{equation}
Its very simple evolution
\begin{equation}
\label{eq:evolution q nonlineaire} q_{NL}(z)=q_{NL}(z_0)+ \frac {m
(z-z_0)} {p_{0 /\!/ }}
\end{equation}
gives readily the real radius of curvature $R(z)$ and the width
$w(z)$ for any altitude $z$. One thus has, as in the linear case,
a simple spherical-wave solution~\eqref{eq:spherical wave
function}. Indeed, this method allows one to approximate very
efficiently the nonlinear propagation of a wave-function of
initial Gaussian profile. Consider a Gaussian atomic beam of width
$w(z_0)=w_0$ at the waist ($R(z_0)=+\infty$) situated at the
position $z_0$ on the propagation axis. Eqs.~\eqref{eq:nonlinear
q} and~\eqref{eq:evolution q nonlineaire} show that the beam width
follows:
\begin{equation}
\label{eq:width expansion} w(z)=\sqrt{w_0^2+ \frac {\hbar^2}
{w_0^2 p_{0 /\!/ }^2} \sigma (z-z_0)^2}
\end{equation}
The width of a self-interacting atomic beam evolves thus as an
interaction-free beam in which the propagation length from the waist
is multiplied by a factor $\sqrt{\sigma}$. As far as the paraxial
beam width evolution is concerned, self-interaction effects thus
operate as a scaling transformation of the free propagation with a
factor $\sqrt{\sigma}$. The quantity $\sqrt{\sigma}-1$ has the same
sign as the scattering length $a$, so one checks that
Eq.~\eqref{eq:width expansion} leads consistently to a faster
expansion for repulsive interactions and to a slower expansion for
attractive ones. As in optics, this treatment can thus be applied to
discuss the self focusing for matter waves. It is, however,
important to keep in mind its validity domain and the several
hypothesis required - constant longitudinal velocity, cylindrical
symmetry, paraxial propagation and Gaussian shape approximation -.
Last, we point out the independent work of Chen~\textit{et.
al.}~\cite{Chen08} on this nonlinear radius of curvature.\\

\section{Second-order computation of the Nonlinear ABCD matrix.}
\label{app:second order}

\subsection{Expression of the second-order matrix.}

In this Appendix, we discuss the nonlinear corrections to the ABCD
matrix associated with the second-order term of the Magnus expansion
$\Omega_2(t,t_0)=  \frac 1 2 \int_{t_0}^{t} dt_1 \int_{t_0}^{t_1}
dt_2  \left[ N(t_1), N(t_2) \right]$, which reads:
\begin{eqnarray}
\label{eq:expression omega2}
 \Omega_2(t,t_0)  =   \left(
\begin{array}{cc}
S(t,t_0) & 0 \\
0 & - S(t,t_0)  \\
\end{array} \right) \nonumber \\
\mbox{with} \: \: S(t)= \int_{t_0}^{t} dt_1 \int_{t_0}^{t_1} dt_2
(\gamma(t_1) - \gamma(t_2) )
\end{eqnarray}
This term, arising from the non-commutativity between the
Hamiltonians taken at different times, naturally depends on the
ordering chosen for the successive lenses. Because of the cloud
expansion, lenses are ordered from the most divergent to the less
divergent. To discuss the effect of this second-order contribution
on the wave-function, it is useful to compute the exponential:
\begin{equation}
\label{eq:exponentiation deuxieme term Magnus expansion} \exp
[\Omega^{(2)}(t,t_0)]=  \left(
\begin{array}{cc}
e^{S(t,t_0)} & 0  \\
0 &  e^{-S(t,t_0)} \\
\end{array}
 \right)
\end{equation}
 The action of such matrix onto the position-momentum width vector $(X,Y)$, defined in Sec. \ref{eq:ABCD theorem}, would
 operate a squeezing between position and momentum. This squeezing is indeed a consequence of our aberrationless approximation, in which the propagation leaves the phase-space volume
  invariant: the expansion of the cloud size must be, in our treatment, compensated
  by a reduced momentum dispersion. One finds consistently that the diagonal matrix elements  $S_{xx,yy,zz}(t)$,
involved in~\eqref{eq:exponentiation deuxieme term Magnus
expansion},
  are positive, which results from the decrease of the matrix elements $\gamma_{xx,yy,zz}(t)$ with time.\\

  The ABCD matrix obtained from a second-order approximation of the Magnus expansion reads:
  \begin{widetext}
\begin{equation}
\label{eq:second order ABCD matrix} M^{(1)}_2(t,t_0,X_0) \simeq
\otimes_{i=x,y,z} \left(
\begin{array} {cc}
  \cosh  K_{ii}(t,t_0)+ S_{ii}(t,t_0)  \frac {\sinh  K_{ii}(t,t_0)} {K_{ii}(t,t_0)} &
(t-t_0)  \frac {\sinh  K_{ii}(t,t_0)} {K_{ii}(t,t_0)} \\
{\langle\gamma \rangle} \frac {\sinh  K_{ii}(t,t_0)}
{K_{ii}(t,t_0)} & \cosh K_{ii}(t,t_0)- S_{ii}(t,t_0)  \frac {\sinh
K_{ii}(t,t_0)} {K_{ii}(t,t_0)}
\end{array}
 \right)
\end{equation}
We have introduced the functions $K_{ii}(t,t_0)=
\sqrt{S_{ii}^2(t,t_0)\: + \: \langle\gamma_{ii} \rangle \:
(t-t_0)^2}$. An analytic expression of $\langle\gamma_{ii} \rangle $
 can be found  for cigar-shaped condensates in
Eq.~\eqref{eq:calcul gamma cigar condensate} of
Appendix~\ref{app:cigar shaped}. The computation of the quantity
$S_{ii}(t,t_0)$ is straightforward, but it involves tedious algebra.
Higher-order contributions to the ABCD matrix~\eqref{eq:expression
martix M Magnus expansion} involve various integrations which need
to be performed numerically.\\

\subsection{Comparison with the first-order matrix.}
Let us expand the matrix~\eqref{eq:second order ABCD matrix} in
the short duration limit. We consider an atomic cloud initially
described by a Gaussian wave-function~\eqref{eq:Gaussian wave
function} of spherical symmetry i.e. $w_{x 0} =w_{y 0} =w_{z 0}
=w_0$. Such assumption does not change the nature of the
discussion, but it considerably simplifies the algebra: the $3
\times 3$ matrices $\gamma(t)$, $S(t)$ and $K(t)$ are then
proportional to the matrix identity $I_{3}$ and can be identified
to scalars. $\gamma(t)$ can be expressed as a function of two time
scales $\tau_1,\tau_2$ involving the sample radius $w_0$, the
scattering length $a$ and fundamental constants:
\begin{equation}
\gamma(t)=  \tau_2^{-2} \left(1+ \frac {(t-t_0)^2} {\tau_1^2}
\right )^{-5/2}, \: \tau_1= \frac {m w^2_0} {\hbar}, \: \tau_2=
\sqrt{\frac {w_0} {4 \pi a}} \tau_1
\end{equation}
 The quantity
$S(t)$~\eqref{eq:expression omega2} can be expressed thanks to a
second-order Taylor expansion of $\gamma(t)$. Setting $\tau=t-t_0$
and noticing that $\gamma'(t_0)=0$, one obtains:
\begin{equation}
S(t)=-\frac {5} {6} \frac {\tau^4} {\tau_1^2 \tau_2^2} + O\left(
\tau^6 \right)
\end{equation}
which yields for the quantity $K(t)$:
\begin{equation}
K(t)=  \sqrt{\langle \gamma \rangle} \tau  \left( 1+ \frac {25}
{72} \frac { \tau^6} {\tau_1^4 \tau_2^2} \right)+ O\left(\tau^8
\right)
\end{equation}
Using this expansion and that of $x\rightarrow \sinh x/ x$, one can
express the second-order matrix $M_2^{(1)}(\tau,X_0)$ as:
 \small
\begin{eqnarray}
\label{eq:second order ABCD matrix} M_2^{(1)}(\tau,X_0) =
M_1^{(1)}(\tau,X_0)+
 \left(
\begin{array}{cc}
- \frac {5} {6} \frac {\tau^4} {\tau_1^2 \tau_2^2} \frac {\sinh
\left( \langle\gamma \rangle^{1/2} \tau \right)}  { \langle\gamma
\rangle^{1/2} \tau}  & \frac {25} {72} \frac { \tau^6} {\tau_1^4
\tau_2^2} \left( \cosh (\langle \gamma \rangle^{1/2} \tau)
- \frac {\sinh \left( \langle \gamma \rangle^{1/2}  \tau \right)} {\langle \gamma \rangle^{1/2}  \tau}  \right)  \\
\frac {25} {72} \frac { \tau^6} {\tau_1^4 \tau_2^2} \left( \cosh
(\langle \gamma \rangle^{1/2}  \tau) - \frac {\sinh \left( \langle
\gamma \rangle^{1/2}  \tau \right)} {\langle \gamma \rangle^{1/2}
\tau}  \right)  & \frac {5} {6} \frac {\tau^4} {\tau_1^2 \tau_2^2}
\frac {\sinh \left( \langle\gamma \rangle^{1/2} \tau \right)} {
\langle\gamma \rangle^{1/2} \tau }    \\
\end{array} \right) + O( \tau^8 )
\end{eqnarray}
\normalsize This expansion shows that the first-order term is a
valid approximation as long as:
\begin{equation}
\tau \ll \tau_c= \left( \frac {w_0} {4 \pi a} \right)^{1/6} \frac
{m w_0^{2}} { \hbar}
\end{equation}
Considering for an instance an initial cloud size of $w_0= 25 \: \mu
\mbox{m}$ and the $^{87}$Rb scattering length $a=5,7 nm$, one
obtains $\tau_1 = 0,14 \: \mbox{s}$, $\tau_2 = 1,63 \: \mbox{s}$,
and $\tau_c=0,31 s$. Note that the relevant small parameter
$\epsilon$, weighting the relative correction brought by the
second-order term, decreases as $\epsilon= (\tau/\tau_c)^{4}$ when
$\tau/\tau_c \rightarrow 0 $.\\

\section{ABCD MATRIX ELEMENTS FOR THE CIGAR-SHAPED CONDENSATE}
\label{app:cigar shaped}

We evaluate in this appendix various primitives necessary to
explicit the non-linear ABCD matrix to first order in the Magnus
expansion given in Eq.~\eqref{eq:first order ABCD matrix}. We
consider a cigar-shaped cylindrical condensate with a long vertical
extension: $w_x=w_y=w_r \ll w_z$. We remind the linear evolution of
the width given by Eq.~\eqref{eq:linear width evolution} i.e.
$w_{r,z \: t}=\sqrt{w_{r,z 0}^2 + \Delta v^2_{r,z 0} (t-t_0)^2}$. We
use the short-hand notation $\Delta
v_{r,z 0}= \hbar /(m w_{r,z 0})$.\\

We seek to evaluate the average $\langle \gamma \rangle_{ii}=
1/\tau \int_{t_0}^t dt \gamma_{ii} (t)$ of the time-dependent
coefficients:
\begin{equation}
\gamma_{rr}(t)= \frac {\gamma_{0}} {w_{z t} w_{r t}^4}, \:
\gamma_{zz}(t)= \frac {\gamma_{0}} {w_{z t}^{3} w_{r t}^2}
\nonumber
\end{equation}
with $\gamma_{0}= g_{I} / ((2 \pi)^{3/2} m)$. These quantities are
readily obtained:
\begin{eqnarray}
\label{eq:calcul gamma cigar condensate} \langle \gamma_{rr}
\rangle & = & \gamma_0 \left[ \frac {\Delta v_{r 0}^2 \: w_{z t}}
{ 2 w_{r 0}^2 ( \Delta v_{r 0}^2 w_{z 0}^2 - w_{r 0}^2 \Delta v_{z
0}^2) w_{r t}^2} + \frac {(\Delta v_{r 0}^2 w_{z 0}^2 - 2 w_{r
0}^2 \Delta v_{z 0}^2) \: \mbox{Arctan} \: \Lambda(t)} {2 w_{r
0}^{3} ( \Delta
v_{r 0}^2 w_{z 0}^2-w_{r 0}^2 \Delta v_{z 0}^2) ^{3/2}  (t-t_0)} \right] \nonumber \\
\langle \gamma_{zz} \rangle & = & \gamma_0 \left[ \frac {\Delta
v_{z 0}^2} { w_{z 0}^2 (\Delta v_{z 0}^2 w_{r 0}^2 - \Delta v_{r
0}^2 w_{z 0}^2 ) w_{z t}} + \frac {\Delta v_{r 0}^2 \:
\mbox{Arctan} \: \Lambda(t)} {w_{r 0} ( \Delta v_{r 0}^2 w_{z
0}^2-w_{r 0}^2 \Delta v_{z 0}^2) ^{3/2}(t-t_0)} \right] \nonumber
\\
 & \: & \mbox{with} \quad \Lambda(t)=\frac {\sqrt{\Delta v_{r
0}^2 w_{z 0}^2 - w_{r 0}^2 \Delta v_{z 0}^2} (t-t_0)} {w_{r t}
w_{z t}}
\end{eqnarray}
\end{widetext}

\end{document}